\documentclass[%
reprint,
amsmath,amssymb,
aps,pre
]{revtex4-2}
\usepackage{graphicx}% Include figure files
\usepackage{bm}
\usepackage{xcolor}% bold math
\begin{document}

	\title{
		%Option 1: First Principles Fourth Order Field Theory for hard-core  Active Brownian Particles\\
		Microscopic derivation  of a field equation for active Brownian particles
		%Option 3: Microscopic derivation  of the active model B+ equations for active Brownian particles\\
	}% Force line breaks with \\
	
	\author{Martín Pinto-Goldberg}
	\affiliation{Departamento de Física, Facultad de Ciencias Físicas y Matemáticas, Universidad de Chile, Avenida Blanco Encalada 2008, Santiago, Chile}
	\author{Rodrigo Soto}
	\affiliation{Departamento de Física, Facultad de Ciencias Físicas y Matemáticas, Universidad de Chile, Avenida Blanco Encalada 2008, Santiago, Chile}
	
	\date{\today}
	
	\begin{abstract}
		
		To understand the phenomena displayed in active phase separation, general top-down theories like Active Model B+ (AMB+) add fluxes that break time reversal symmetry. Starting from an Enskog-like kinetic theory of hard-core active Brownian particles in the high persistence regime, we derive AMB+ from first principles. For the effective free energy to have two minima, we propose an effective parametrization of the pair correlation function. Explicit expressions for all coefficients in the model are given as a function of the microscopic parameters to leading order in the Péclet number.
	\end{abstract}
	
	\maketitle

	{\it Introduction}.--The individual units that compose an active system are able to self-propel by constantly taking energy from their surroundings or internal reservoirs~\cite{RevModPhys.88.045006,annurev:/content/journals/10.1146/annurev-conmatphys-070909-104101,shaebani2020computational}. This permanent flux of energy implies that active systems are inherently out of thermodynamic equilibrium and can show a rich variety of collective behavior. Instances arise in systems of living cells \cite{Carrre2023, Hannezo2022, Butcher2009,NORDENTOFT2026100574,Lenormand2008}, suspensions of self-propelled colloidal particles \cite{PhysRevLett.110.238301, annurev:/content/journals/10.1146/annurev-conmatphys-031214-014710}, biological microswimmers~\cite{B812146J,PhysRevLett.122.248102,Kearns2010}, among others. 
	
	Notably, active phase separation can occur in systems of purely repulsive self-propelled particles. Microscopically, steric interactions cause noninertial persistent active particles to slide against each other until the moment of detachment. While they  are in contact, others can arrive, starting the formation of a cluster ~\cite{PhysRevLett.108.235702,PhysRevLett.110.055701}. This implies that particles will move more slowly  in regions of higher densities and will accumulate there \cite{annurev:/content/journals/10.1146/annurev-conmatphys-031214-014710,Cates2013}. This accumulation can lead to complete phase separation between dense and  dilute regions \cite{PhysRevLett.108.235702,PhysRevLett.110.055701,PhysRevLett.110.238301}, or to microphase separated states, comprising dense clusters in a sea of vapor or dilute bubbles in a liquid \cite{PhysRevLett.108.268303,10.1039/c3sm52813h,PhysRevLett.125.168001}. From a phenomenological perspective, field theories have been proposed to study scalar phase separation in active systems, where the particle density is the only conserved field. These theories stem from model B of the Halperin and Hohenberg classification \cite{RevModPhys.49.435}, which describes systems undergoing equilibrium diffusive fluid-fluid phase separation. Then, with model B as the basis, activity is introduced in the form of nonequilibrium fluxes which break time reversal symmetry, meaning that they can not be derived from a free energy functional \cite{Meissner2024}.
	
	Active Model B+ (AMB+) \cite{PhysRevX.8.031080} is one of these theories, describing the conserved dynamics of a scalar field $\phi(\mathbf{r},t)$ which is a linear transform of the particle density $\rho(\mathbf{r},t)$, such that $\phi=0$ at the critical density. In AMB+ the deterministic dynamical equations are
	\begin{align}\label{eq:AMB+} 
		\frac{\partial \phi}{\partial t}&=-\nabla\cdot\mathbf{J}\\
		\mathbf{J}&=-\nabla\left[\mu_{\text{eq}}+\lambda(\nabla\phi)^2\right]+\zeta (\nabla^2\phi)\nabla\phi,\label{eq:AMBJ+} 
	\end{align}
	where $\mathbf{J}$ is the rescaled density flux and $\mu_\text{eq}=\delta F/\delta \phi$ is the equilibrium chemical potential deriving from a free energy of the form
	\begin{equation}\label{eq:F}
		F[\phi]=\int \left[g(\phi)+\frac{K}{2}(\nabla\phi)^2\right]d\mathbf{r}.
	\end{equation}
	Here, $g(\phi)$ is the bulk free energy density which is usually taken to be of a $\phi^4$ form, with two energy minima representing the coexisting densities. The nonequilibrium transport coefficients $\lambda$ and $\zeta$, with their respective fluxes, break time-reversal symmetry and effectively drive the system out of equilibrium as they cannot be derived from a free energy. The dynamics of AMB+ have been shown to reproduce a variety of states for the different regions of the $\lambda$-$\zeta$ parameter space~\cite{PhysRevX.8.031080, Cates_2025,PhysRevLett.127.068001}. These include standard bulk phase separation and when Ostwald ripening is reversed, bubbly and microphase separation.
	
	It is still not fully understood what microscopic conditions are necessary for the different phase separated states to emerge. As such, deriving continuum theories from microscopic models is a necessary task to relate the underlying microscopic dynamics to the observed macroscopic phenomenology.
	Active Brownian particles (ABP) constitute one of the simplest models that present active scalar phase separation. These are spherical self-propelled particles of diameter $\sigma$ that move persistently at constant speed $V$, in a direction that changes diffusively at rate $D_r$. 
	Coarse-graining particle models like ABP into continuum theories always poses a challenge due to particle interactions. One way to encapsulate them is the quorum sensing (QS) model \cite{annurev:/content/journals/10.1146/annurev-conmatphys-031214-014710,Cates2013}, which assumes a mean-field density dependent particle velocity. Starting from the Langevin equation for QS particles and then eliminating the fast orientation variables, Dean's equation \cite{dean1996langevin} can give a theory of the same order of AMB+, although with no counterpart to the $\zeta$ term in the latter \cite{Cates_2025}. Employing a multiple scale analysis results in a theory that allows to study the system dynamics beyond weak phase separation \cite{burekovic2026activecahnhilliardtheorynonequilibrium}. In the case of self propelled particles interacting with soft repulsive potentials, it is also possible to obtain fourth order field theories for the particle density with methods akin to dynamical density functional theory, where varying degrees of approximation are applied \cite{PhysRevResearch.2.033241,Vrugt_2023,Bickmann_2020,doi:10.1073/pnas.2219900120,PhysRevE.110.054604}. In these cases, the nonequilibrium parameters depend on the choice of interparticle potential. 
	
	The case of hard-core, noninertial ABP presents a challenge in the derivation of coarse-grained descriptions due to the difficulty in characterizing steric interactions. A geometric approach, developed by Bruna and Chapman \cite{PhysRevE.85.011103,10.1063/1.4767058}, consists of restricting the domain of the evolution equations for the $N$-particle density in areas where particles overlap. This was successfully applied to derive macroscopic equations for hard-core ABP~\cite{doi:10.1137/21M1452524} and later used to derive a theory like AMB+ for the active chiral particle model \cite{Kalz_2024}.  In this article, we follow a kinetic description of an ensemble of hard-core ABP to present a bottom-up derivation of a fourth order field theory, which has the same structure as the deterministic part of AMB+. This implies that we obtain explicit expressions for all coefficients in terms of the microscopic parameters.
	
\begin{figure}[h]
		\centering
		\includegraphics[width=\linewidth]{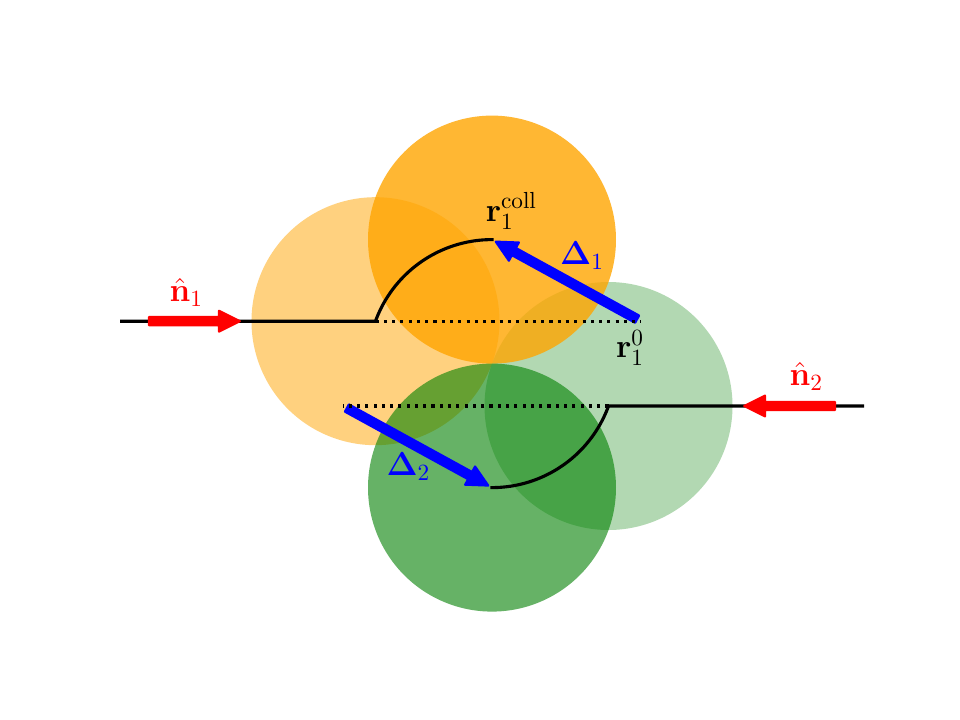}
		\caption{Collision scheme for ABP interaction in the limit of infinite persistence. For simplicity we illustrate the case of $\mathbf{\hat{n}}_2=-\mathbf{\hat{n}}_1$. Particle directors are shown in red. The solid black lines represent the trajectories of the particles during the collision, while the dotted lines show the particle motion in the absence of it. The end points of these trajectories are $\mathbf{r}_i^{\rm{coll}}$ and $\mathbf{r}_i^{0}$, respectively. The effective displacements $\mathbf{\Delta}_i$ caused by the interaction are in blue.}
		\label{fig:coll}
	\end{figure}

	{\it Kinetic theory for active Brownian particles}.--The fully microscopic derivation of the deterministic part of AMB+ presented here follows from a recently proposed kinetic description of ABP. 
	Here, we consider particles moving in two dimensions, where each particle is described by the equations of motion
	\begin{equation}\label{eq:ABP}
		\dot{\mathbf{r}}=V\hat{\mathbf{n}}+\mathbf{F},\quad \dot{\theta} = \sqrt{2D_r}\xi(t),
	\end{equation}
	with $\hat{\mathbf{n}}=(\text{cos}~\theta,\text{sin}~\theta)$, $\mathbf{F}$ a force that imposes excluded volume with the rest of the particles, and $\xi(t)$ a Gaussian white noise term. A kinetic description of an ensemble of ABP is not straightforward due to the noninertial nature of their interactions~\cite{mayo2026cooling}. However, it was recently shown in Ref.~\cite{PhysRevLett.132.208301} that in the high persistence regime, characterized by large P\'eclet number $\text{Pe}\equiv V/(\sigma D_r)\gg1$, the interactions can be modeled as instantaneous events changing the positions of the particles, but not their directors, which are assumed to remain constant for the duration of the collision. Following the schematic of Fig.~\ref{fig:coll}, the effect of collisions is to displace particles by $\bm{\Delta}_i = \mathbf{r}_i^\text{coll}-\mathbf{r}_i^0$ at the moment both particles come into contact, where $\mathbf{r}_i^\text{coll}$ is the actual final position of the particles following Eqs.~\eqref{eq:ABP} and $\mathbf{r}_i^0$ would be to final positions in absence of the collision. 
	
	With the previous description, the ensemble of ABP can be described by the distribution function $f(\mathbf{r}_1,\hat{\mathbf{n}}_1,t)$ whose evolution is given by \begin{equation}\label{eq:kin}
		\frac{\partial f}{\partial t} + V\mathbf{\hat{n}_1}\cdot\frac{\partial f}{\partial\mathbf{r}_1} = D_r\frac{\partial^2 f}{\partial \theta_1^2}+ J[f,f].
	\end{equation}
	Here the first three terms account for the free streaming of particles and rotational diffusion, while $J[f,f]$ is the collision operator. Following the description of interactions given previously, the collision term is 
	\begin{multline}\label{eq:J}
		J[f,f] = \int \chi\left(\mathbf{r}'_1+\frac{\sigma\hat{\bm{\sigma}}}{2}\right)f(\mathbf{r}'_1,\mathbf{\hat{n}}_1) f(\mathbf{r}'_1+\sigma\hat{\bm{\sigma}},\mathbf{\hat{n}}_2)
		\\\times |V\sigma (\mathbf{\hat{n}}_2-\mathbf{\hat{n}}_1)\cdot\hat{\bm{\sigma}}|\Theta[-(\mathbf{\hat{n}}_2-\mathbf{\hat{n}}_1)\cdot\hat{\bm{\sigma}}]\\\times[\delta(\mathbf{r}_1-\mathbf{r}_1^\prime-\bm{\Delta}_1) -\delta(\mathbf{r}_1-\mathbf{r}^\prime_1)]d\mathbf{r}_1^\prime d\mathbf{\hat{n}}_2d\hat{\bm{\sigma}},
	\end{multline}
	where $\chi$ is the pair correlation function evaluated at the point of contact, $\hat{\bm{\sigma}}$ is the unit vector pointing from particle 1 to particle 2 and $\Theta$ is the Heaviside function which selects only collisional trajectories., The Dirac deltas enforce the particle is being gained and lost at $\mathbf{r}_1$, needing for one colliding particle to to be at $\mathbf{r}_1-\bm{\Delta}_1$ and $\mathbf{r}_1$, respectively, while the partner is displaced by $\sigma\hat{\bm{\sigma}}$

	{\it Active model B+}.--Because we are interested in deriving AMB+, first we need to obtain an equation at least to fourth order in gradients for the particle density, defined as the first moment of the distribution function over the director $\hat{\mathbf{n}}$
	\begin{equation}\label{eq:rhoA}
		\rho(\mathbf{r}_1,t)\equiv \int f(\mathbf{r}_1,\mathbf{\hat{n}_1},t)d\mathbf{\hat{n}}_1.
	\end{equation}
	Finally, the derived equation must be expanded close to the spinodal density,  $\rho(\mathbf{r}_1,t)=\phi(\mathbf{r}_1,t)+\rho_s(1+\epsilon)$, keeping all terms of order $\mathcal{O}(\nabla^2\phi^4$) and $\mathcal{O}(\nabla^4\phi^2$). Here $\rho_s$ denotes the spinodal density and $\epsilon$ is the parameter that tells us how far we are from the former. 
	
	The kinetic equation allows obtaining hydrodynamic equations by taking its appropriate moments. However, doing this directly does not give a closed equation for $\rho(\mathbf{r}_1,t)$ because an unknown flux appears which depends non locally on the distribution function itself.  To obtain a closed equation we  use the Chapman--Enskog method, which consists of a multiscale method that systematically solves the kinetic equation via a gradient expansion \cite{chapman1990mathematical,Cercignani1988}. In a previous work we obtained a set of diffusive order equations for the density and polarization fields for ABP \cite{fg3f-763d}. While it is possible to obtain a fourth order equation for the density by adiabatically eliminating the polarization, the resulting equation would be missing some terms. This is because integrating out the fast mode and performing the gradient expansion of the hydrodynamic equations are operations that do not commute. 
	
	The Chapman--Enskog solution of the kinetic equation is outlined here, while a more detailed derivation is presented in the Supplemental Material. First we must assume that the kinetic equation accepts normal solutions, meaning that the spatiotemporal dependence is enslaved to the conserved field $\rho$. Then, the distribution function is expanded in powers of a small formal parameter $\varepsilon$, that is proportional to each spatial gradient, $f=f^{(0)}+\varepsilon f^{(1)}+\varepsilon^2f^{(2)}+\dots$. Lastly, timescale separation is made explicit by the introduction of several time variables $t_0=t$, $t_1=\varepsilon t$, $t_2=\varepsilon^2 t$, and so forth. Using these assumptions the distribution function can be written as $f(\mathbf{r}_1,\hat{\mathbf{n}}_1,t)=f[\rho(\mathbf{r}_1,t_0,t_1,t_2,\dots);\hat{\mathbf{n}}_1]$ and due to the chain rule, the time derivative will give terms in different powers of $\varepsilon$. To calculate the Chapman--Enskog solution it is also necessary to expand the collision operator~(\ref{eq:J}) in gradients because of its non local form. As shown in the Supplemental Material this gives an expansion of the form $J = \varepsilon J^{(1)}+\varepsilon^2J^{(2)}+\dots$.

	The introduction of all the previously discussed elements in the kinetic equation gives a hierarchy of equations which can be solved in increasing orders of $\varepsilon$. At zeroth order in $\varepsilon$ one obtains that the density remains constant at the $t_0$ scale, which must hold due to particle conservation. Then, solving for $f^{(0)}$ gives
	\begin{equation}
		f^{(0)} =\frac{\rho}{2\pi},
	\end{equation}
	where we imposed the normalization condition that the moment of $f^{(0)}$ must reproduce the particle density exactly. 
	
	By virtue of the conservation of the density and because it is a scalar field, at order $\varepsilon^1$ it is also found that $\partial_{t_1}\rho=0$. Then, using that $\partial_{\theta_1\theta_1}\hat{\mathbf{n}}_1=-\hat{\mathbf{n}}_1$ it is easily found that
	\begin{equation}\label{eq:f1}
		f^{(1)} = -\frac{\hat{\mathbf{n}}}{2\pi D_r}\cdot\nabla[v(\rho)\rho]
	\end{equation}
	with $v(\rho)=V[1-\pi\sigma^2\chi(\rho)\rho/4]$, the effective density dependent velocity due to displacements.
	
	At second order in gradients,  the first non-trivial hydrodynamic equation appears. Taking the moment of the kinetic equation yields the diffusion equation
	\begin{equation}\label{eq:rho2}
		\frac{\partial \rho}{\partial t_2}  = \nabla\cdot[D(\rho)\nabla\rho ],
	\end{equation}
	where
	\begin{equation}
		D(\rho)=D_0\left[1-\frac{(\rho^2\chi)'}{2\rho^*}  \right]
	\end{equation}
	with $D_0 = V^2/(2D_r)$ the bare diffusion coefficient for ABP,
	\begin{equation}
		\rho^*=\frac{2/(\pi\sigma^2)}{1-\frac{16C}{\pi^2\text{Pe}}}
	\end{equation}
	the spinodal density in the absence of correlations ($\chi=1$), and $C\approx0.916$ the Catalan constant.  Hereafter, the primes denote derivatives with respect to the density. 
	
	To study the density instability that causes phase separation in ABP one could usually stop at the diffusive order, but since we are interested an equation like AMB+, we must solve for $f^{(2)}$ and keep going until order $\varepsilon^4$. In the Supplemental Material it is explained how $f^{(2)}$ is solved and its solution results in a functional that is proportional to second derivatives of the density, whose explicit form we do not show here for brevity, contracted with the tensor product of $\mathbf{\hat{n}}$ with itself, $n_{1\alpha}n_{1\beta}$.
	
	Similar to the first order kinetic equation, at third order in $\varepsilon$ we have $\partial_{t_3}\rho=0$, because the density is a scalar quantity and we can not form one with three gradients. The solution for $f^{(3)}$ is explained in the Supplemental Material and it contains third derivatives of the density, contracted with $n_{1\alpha}n_{1\beta}n_{1\gamma}$. Once again, the explicit expressions are long and contain many coefficients, thus they are not shown here. Taking the moment of the fourth order kinetic equation, our stopping point, allows us to obtain an equation for the density which depends on four derivatives and with all coefficients  depending on the microscopic parameters. Finally, the dynamics at all orders are summed by replacing $t_n=\varepsilon^nt$ and then $\varepsilon$ is set to unity. The resulting equation has the form
	\begin{equation}\label{eq:rhofull}
		\frac{\partial \rho}{\partial t} = \nabla\cdot\{D(\rho)\nabla\rho+\mathbf{G}[\rho]\},
	\end{equation}
	where  $\mathbf{G}[\rho]$ is of order $\nabla^4$ and all of its terms are given in the Supplemental Material due to their length. 
	
	From this equation, we linearize around the spinodal density, which must satisfy $D(\rho_s)=0$. This condition translates to 
	\begin{equation}\label{eq:rhos}
		\rho_s [\chi(\rho_s)+\rho_s\chi'(\rho_s)/2]=\rho^*.
	\end{equation}
	Next, we proceed by expanding close to $\rho_s$ and then keeping all terms of order $\mathcal{O}(\nabla^2\phi^4)$ and $\mathcal{O}(\nabla^4\phi^2)$. Furthermore, we keep terms of order $\epsilon$.
	The resulting equation for $\phi$ has the AMB+ form of Eq.~(\ref{eq:AMB+}), including both nonequilibrium fluxes in \eqref{eq:AMBJ+}. The associated free energy density is
	\begin{equation}\label{eq:f}
		g(\phi)=\epsilon \frac{D'(\rho_s)\rho_s}{2}\phi^2+\frac{D'(\rho_s)}{6}\phi^3+\frac{D''(\rho_s)}{24}\phi^4
	\end{equation}
	and the Ginzbug coefficient $K$ depends on the field, 
	\begin{equation}
		K(\phi)=\kappa-\xi\phi.
	\end{equation}
	
	All parameters in the field equations $\kappa$, $\xi$, $\lambda$ and $\zeta$ depend explicitly on the P\'eclet number and implicitly on the pair correlation function $\chi$.
	To give explicit expressions for these coefficients, we must make a few remarks. Because the kinetic theory proposed above is valid in the high persistence regime, for simplicity from here onward all results consider the leading contribution in the Péclet number for $\text{Pe}\gg1$. This means, for example, that $\rho^*\approx2/(\pi\sigma^2)$ and then the spinodal density is the same as the one found in Ref.~\cite{PhysRevLett.132.208301}. Next, we need a form for the pair correlation function $\chi$. First, we have the choice to neglect correlations by letting $\chi=1$, meaning that we consider a Boltzmann-like collision operator. To consider correlations, as in the Enskog theory for moderately dense gases \cite{10.1093/acprof:oso/9780198716051.001.0001}, due to the lack of a pair correlation function at contact specific for ABP, we can take it to be that of hard disks in equilibrium \cite{Henderson01091975} $\chi_{\text{hd}}=(1-7\pi\sigma^2\rho/64)/(1-\pi\sigma^2\rho/4)^2$. Both of these choices give qualitatively the same results with non vanishing explicit expressions for the nonequilibrium coefficients and carry the same caveats. 
	The first term in the free energy density in Eq.~(\ref{eq:f}) dictates the stability of the homogeneous state and in both the Boltzmann and the Enskog case we have $D'(\rho_s)<0$, meaning that the instability is correctly encoded in $g(\phi)$. Then, to get nonlinear saturation and two energy minima, representing the binodals, we require $D''(\rho_s)>0$. This condition, however, is not satisfied with either choice of $\chi$. In the Boltzmann case, the second and all higher derivatives of $D(\rho)$ simply vanish and in the Enskog case, all its derivatives are negative.

	As a third option, we propose an effective parametrization of the pair correlation function. For the parametrization to be reasonable, we require that it satisfies four conditions. First, in the low density limit, correlations are absent $\chi(\rho\rightarrow0)=1$. Second, we require that $\chi$ is a monotonically increasing function of density for all densities in the range $[0,\rho_\text{max}]$, with $\rho_\text{max}\sigma^2=2/\sqrt{3}$ the close packing density. Then, the spinodal density is asked to be close to $\rho_s\sigma^2\approx 0.32$, which matches simulations of ABP with hard disk interactions for infinitely large Pe \cite{PhysRevResearch.2.023010,PhysRevLett.121.098003,10.1039/c7sm01504f}. Finally, to have saturation we need $D''(\rho_s)>0$.  
	Several functional forms satisfy these conditions. For simplicity and not suggesting this to be the real form that should be obtained from simulations, we propose an effective correlation function in the form $\chi_\text{eff}(\rho)=1+A\rho-B\rho^2+C\rho^3$, with $A=3.14\sigma^2$, $B=3.7\sigma^4$ and $C=3.1\sigma^6$. With this choice, the numerical values for the coefficients to leading order in Pe are
	\begin{align}
		&\kappa\approx-(0.0038\epsilon \text{Pe}^3+0.0056\text{Pe}^2)V\sigma^3,\\
		&\xi\approx0.0087V\sigma^5\text{Pe}^3\\
		&\lambda\approx 0.0044V\sigma^5\text{Pe}^3\\
		&\zeta\approx-(0.048\epsilon \text{Pe}^3+0.070\text{Pe}^2)V\sigma^5.
	\end{align}
	In these expressions, specifically for $\kappa$ and $\zeta$, we show the first terms despite being of higher order in $\epsilon$. With this we highlight that the leading contributions in Pe for both of these coefficients change sign at the spinodal density. 
	
	The nonequilibrium coefficients $\lambda$ and $\zeta$ grow in absolute value rapidly with the Péclet number, which is usually used as a measure of the activity of the system. Furthermore, these coefficients have opposite signs, meaning that according to phase diagrams for AMB+ \cite{PhysRevX.8.031080,Cates_2025}, they predict coarsening and complete phase separation for ABP.
	Although an effective correlation function is needed to correctly account for the nonlinear saturation, when using the hard disk expression for $\chi$, the numerical values of the coefficients are close as shown in the End Matter. This might be related to the fact that these coefficients depend on the values of $\chi$ close to $\rho_s$, quite independent on its behavior at close packing.
	
	The parameter $\kappa$ that quantifies the energy cost of spatial inhomogeneities is negative, which would result in the emergence of infinitely rough states beyond $\rho_s$. However, the numerical analysis of the kinetic equation in  Ref.~\cite{PhysRevLett.132.208301} showed that the small wavelength modes are stable, and only the large wavelength modes present the instability. This discrepancy is originated from the fact that the present results come from the Chapman--Enskog expansion only up to fourth order in gradients. Higher order terms are expected to account for the stability of the small wavelength modes.

	{\it Discussion}.--Much more work still remains to be done to understand the microscopic conditions needed for the distinct phenomena observed in active phase separation. Understanding the macroscopic transport properties of specific systems in terms of the microscopic parameters allows to tune the dynamics at the component level to achieve specific behaviors at the macroscale. This is specially important in active matter because of its interdisciplinary nature and potential applications \cite{Gompper2020}. 
	
	ABP as one of the simplest models for noninertial scalar active matter, still presents a challenge in working on fully microscopic derivations of models like AMB+ because steric interactions are difficult to characterize. The effective displacement theory is restricted for high persistence and the effect of reorientantions during collisions could potentially play a role in giving rise to complex structures during coarsening and microphase separation. Furthermore, it was necessary to propose a parametrization for the pair correlation function to ensure nonlinear saturation, otherwise this would lead to a non physical situation. We argue that for the case of noninertial particles, the pair correlation function should not diverge near close packing as opposed to $\chi_\text{hd}$. This is because, contrary to elastic collisions, the collision frequency of noninertial particles at moderate densities does not diverge. 
	
	On the other hand, this kinetic theory does not predict noise terms, which play a crucial role in the dynamic states reproduced in AMB+ \cite{PhysRevX.8.031080,Cates_2025}. A fully microscopic derivation of AMB+ for the ABP model then would require introducing noise, for example in the collision operator. In the present formulation, $J[f,f]$ is associated to the average number of collisions at position $\mathbf{r}$ per unit time. Then, the deviations around this number present a source of stochasticity.

	\bibliography{apssamp}
	
	\clearpage
	%\newpage
	
	{\Large \bf End Matter}
	
	The transport coefficients for the Boltzmann approximation (\textit{i.e.}, $\chi=1$) are
	\begin{align}
		&\kappa\approx-(0.0028\epsilon \text{Pe}^3+0.0049\text{Pe}^2)V\sigma^3,\\
		&\xi\approx0.0045V\sigma^5\text{Pe}^3,\\
		&\lambda\approx 0.0022V\sigma^5\text{Pe}^3,\\
		&\zeta\approx-(0.018\epsilon \text{Pe}^3+0.031\text{Pe}^2)V\sigma^5.
	\end{align}
	
	In the case of using the hard disk correlation function $\chi=\chi_\text{hd}$,
	\begin{align}
		&\kappa\approx-(0.0045\epsilon \text{Pe}^3+0.0060\text{Pe}^2)V\sigma^3,\\
		&\xi\approx0.0082V\sigma^5\text{Pe}^3,\\
		&\lambda\approx 0.0041V\sigma^5\text{Pe}^3,\\
		&\zeta\approx-(0.057\epsilon \text{Pe}^3+0.075\text{Pe}^2)V\sigma^5.
	\end{align}
	
\newpage	
\appendix
\onecolumngrid
\section{Supplemental Material}
\subsection{Gradient expansion of the collision operator}
The derivation of an equation for the particle density starts by expanding the collision operator
\begin{multline}\label{eq:J}
J[f,f] = \int \chi(\mathbf{r}'_1+\sigma\bm{\hat{\sigma}}/2)f(\mathbf{r}'_1,\mathbf{\hat{n}}_1) f(\mathbf{r}'_1+\sigma\hat{\bm{\sigma}},\mathbf{\hat{n}}_2)
|V\sigma (\mathbf{\hat{n}}_2-\mathbf{\hat{n}}_1)\cdot\hat{\bm{\sigma}}|\Theta[-(\mathbf{\hat{n}}_2-\mathbf{\hat{n}}_1)\cdot\hat{\bm{\sigma}}]\\\times[\delta(\mathbf{r}_1-\mathbf{r}_1^\prime-\mathbf{\Delta}_1) -\delta(\mathbf{r}_1-\mathbf{r}^\prime_1)]d\mathbf{r}_1^\prime d\mathbf{\hat{n}}_2d\hat{\bm{\sigma}}
\end{multline}
in gradients, because it has a non-local form. For a field theory such as AMB+ we need to expand up to fourth order in gradients. Using the properties of the Dirac delta function we are left with 
\begin{equation}
J= J^{(1)}+J^{(2)}+J^{(3)}+J^{(4)}+\mathcal{O}(\nabla^5),
\end{equation}
where $J^{(n)}$ is of order $\nabla^n$ and the explicit expressions are
\begin{align}
J^{(1)}(\mathbf{r}_1,\hat{\mathbf{n}}_1) &= -\nabla_\alpha\int \Delta_\alpha\chi f_1 f_2d\mathcal{H} \label{eq:J1},\\
J^{(2)}(\mathbf{r}_1,\hat{\mathbf{n}}_1) &= -\sigma\nabla_\alpha\int \Delta_\alpha\hat{\sigma}_\beta f_1\left(\chi\nabla_\beta f_2+\frac{f_2}{2}\nabla_\beta\chi\right)d\mathcal{H}+\frac{\nabla_\alpha\nabla_\beta}{2}\int \Delta_\alpha\Delta_\beta\chi f_1 f_2d\mathcal{H} \label{eq:J2},\\
J^{(3)}(\mathbf{r}_1,\hat{\mathbf{n}}_1)&= -\frac{\sigma^2}{2}\nabla_\alpha\int\Delta_\alpha\hat{\sigma}_\beta\hat{\sigma}_\gamma f_1\left( \chi\nabla_\beta\nabla_\gamma f_2 +\nabla_\beta\chi\nabla_\gamma f_2+\frac{f_2}{4}\nabla_\beta\nabla_\gamma\chi \right)d\mathcal{H} \nonumber\\
&+\sigma\frac{\nabla_\alpha\nabla_\beta}{2}\int \Delta_\alpha\Delta_\beta\hat{\sigma}_\gamma f_1\left(\chi\nabla_\gamma f_2+\frac{f_2}{2}\nabla_\gamma\chi \right)d\mathcal{H}-\frac{\nabla_\alpha\nabla_\beta\nabla_\gamma}{3!}\int\Delta_\alpha\Delta_\beta\Delta_\gamma \chi f_1 f_2d\mathcal{H} \label{eq:J3},\\
J^{(4)}(\mathbf{r}_1,\hat{\mathbf{n}}_1)&=-\sigma^3\nabla_\alpha\int\Delta_\alpha\hat{\sigma}_\beta\hat{\sigma}_\gamma\hat{\sigma}_\mu f_1\left(\frac{\chi}{3!}\nabla_\beta\nabla_\gamma\nabla_\mu f_2+\frac{1}{4}\nabla_\beta\chi\nabla_\gamma\nabla_\mu f_2+\frac{1}{8}\nabla_\beta\nabla_\gamma\chi\nabla_\mu f_2+\frac{f_2}{8\cdot 3!}\nabla_\beta\nabla_\gamma\nabla_\mu\chi \right)d\mathcal{H} \nonumber\\
&+\sigma^2\frac{\nabla_\alpha\nabla_\beta}{4}\int \Delta_\alpha\Delta_\beta\hat{\sigma}_\gamma\hat{\sigma}_\mu f_1\left(\chi\nabla_\gamma\nabla_\mu f_2+\nabla_\gamma\chi\nabla_\mu f_2+\frac{f_2}{4}\nabla_\gamma\nabla_\mu\chi \right)d\mathcal{H}\nonumber\\
&-\sigma\frac{\nabla_\alpha\nabla_\beta\nabla_\gamma}{3!}\int \Delta_\alpha\Delta_\beta\Delta_\gamma\hat{\sigma}_\mu f_1\left(\chi\nabla_\mu f_2+\frac{f_2}{2}\nabla_\mu\chi \right)d\mathcal{H}+\frac{\nabla_\alpha\nabla_\beta\nabla_\gamma\nabla_\mu}{4!}\int\Delta_\alpha\Delta_\beta\Delta_\gamma\Delta_\mu \chi f_1 f_2d\mathcal{H}. \label{eq:J4}
\end{align}
\twocolumngrid
In the expressions,  all functions are evaluated at $\mathbf{r}_1$, we used the notation 
\begin{equation}
d\mathcal{H}=V\sigma|(\mathbf{\hat{n}_2}-\mathbf{\hat{n}}_1)\cdot\bm{\hat{\sigma}}|\Theta[(\mathbf{\hat{n}}_1-\mathbf{\hat{n}}_2)\cdot\bm{\hat{\sigma}}]d\mathbf{\hat{n}}_2d\bm{\hat{\sigma}},
\end{equation} $f_i=f(\mathbf{\hat{n}}_i)$, and the Einstein summation convention was assumed, with Greek indices indicating Cartesian coordinates.

\subsection{Calculation of the integrals} \label{sect:integrals}
For what follows, it is useful to show the general form of the integrals to be calculated in the derivation of AMB+. At each increasing order in the kinetic equation we will have to solve integrals of the form
\begin{equation}
I_{\alpha\beta\dots\gamma}(\mathbf{\hat{n}}_1) = \int g_{\alpha\beta\dots\gamma}(\mathbf{\hat{n}}_1,\mathbf{\hat{n}}_2,\bm{\hat{\sigma}})d\mathcal{H},
\end{equation}
where the functions $g$ depend on the three unit vectors. Because we will derive an equation just for the density, which is a scalar quantity, we will need the solution up to integrals of the form $I_{\alpha\beta\gamma\mu}$, that is up to four indices.

We start with 
\begin{equation}
I_\alpha(\mathbf{\hat{n}}_1) =\int g_{\alpha}(\mathbf{\hat{n}}_1,\mathbf{\hat{n}}_2,\bm{\hat{\sigma}})d\mathcal{H}.
\end{equation}
Because $\mathbf{\hat{n}}_2$ and $\bm{\hat{\sigma}}$ are integrated out, the only possible solution is a vector that depends on $\mathbf{\hat{n}}_1$. Furthermore, the isotropy of space implies that $I_\alpha(\mathbf{\hat{n}}_1) = a n_{1\alpha}$, with $a$ a constant to be found. 
To obtain $a$ we multiply both sides of the equality by $n_{1\beta}$ and then take the trace $\text{Tr}_{\alpha\beta}$, resulting in $I_\alpha(\mathbf{\hat{n}}_1) n_{1\alpha} = a$, where we used that $\mathbf{\hat{n}}_1$ is a unit vector. Integrating this over $\mathbf{\hat{n}}_1$ gives
\begin{equation}
2\pi a = \int I_{\alpha}(\mathbf{\hat{n}}_1)n_{1\alpha}d\mathbf{\hat{n}}_1.
\end{equation}
With the constant found, we have the solution
\begin{equation}\label{eq:I1}
\begin{split}
	I_\alpha(\mathbf{{\hat{n}}}_1) &= \int g_{\alpha}(\mathbf{\hat{n}}_1,\mathbf{\hat{n}}_2,\bm{\hat{\sigma}})d\mathcal{H} = a n_{1\alpha},\\
	a &= \frac{1}{2\pi}\int g_{\alpha}n_{1\alpha}d\hat{\mathbf{n}}_1d\mathcal{H}.
\end{split}
\end{equation}
For integrals with more indices, the same arguments apply, meaning that we exploit the dependence on $\mathbf{\hat{n}}_1$ and the isotropy of space. For the case of two indices we  have that the result must be a linear combination of all two index tensors that can be found in the system, that is the identity tensor $\delta_{\alpha\beta}$ and the tensor product of $\mathbf{\hat{n}}_1$ with itself $n_{1\alpha}n_{1\beta}$. With this we have
\begin{equation}
I_{\alpha\beta}(\mathbf{\hat{n}}_1) = \int g_{\alpha\beta}(\mathbf{\hat{n}}_1,\mathbf{\hat{n}}_2,\bm{\hat{\sigma}})d\mathcal{H} = a n_{1\alpha}n_{1\beta}+b\delta_{\alpha\beta}.
\end{equation}
Finding $a$ and $b$ requires two equations. We obtain the first one by taking $\text{Tr}_{\alpha\beta}$ on both sides and the second one follows from multiplying both sides by $n_{1\gamma}n_{1\mu}$ and then taking $\text{Tr}_{\alpha\gamma}\text{Tr}_{\beta\mu}$. Furthermore, we integrate out $\mathbf{\hat{n}}_1$ in both equations to obtain
\begin{equation}
a+2b = \frac{1}{2\pi}\int g_{\alpha\alpha}d\hat{\mathbf{n}}_1d\mathcal{H},
\end{equation}
\begin{equation}
a+b = \frac{1}{2\pi}\int g_{\alpha\beta}n_{1\alpha}n_{1\beta}d\mathbf{\hat{n}}_1d\mathcal{H}.
\end{equation}
With both equations it is trivial to obtain an explicit expression for the constants. The final result for this integral is then
\begin{equation}
\begin{split}\label{eq:I2}
	I_{\alpha\beta}(\mathbf{\hat{n}}_1) &= \int g_{\alpha\beta}(\mathbf{\hat{n}}_1,\mathbf{\hat{n}}_2,\bm{\hat{\sigma}})d\mathcal{H} = a n_{1\alpha}n_{1\beta}+b\delta_{\alpha\beta},\\
	a&=\frac{1}{2\pi}\int g_{\alpha\beta}(2n_{1\alpha}n_{1\beta}-\delta_{\alpha\beta})d\hat{\mathbf{n}}_1d\mathcal{H},\\
	b&=\frac{1}{2\pi}\int g_{\alpha\beta}(\delta_{\alpha\beta}-n_{1\alpha}n_{1\beta})d\hat{\mathbf{n}}_1d\mathcal{H}.
\end{split}
\end{equation}
The result for $I_{\alpha\beta\gamma}$ follows the same reasoning and we will just show it
\begin{equation}\label{eq:I3}
\begin{aligned}
	I_{\alpha\beta\gamma}(\mathbf{\hat{n}}_1)
	&= \int g_{\alpha\beta\gamma}(\hat{\mathbf n}_1,\hat{\mathbf n}_2,\hat{\boldsymbol\sigma})\,
	d\mathcal{H}\\
	&= a n_{1\alpha} n_{1\beta} n_{1\gamma}
	+ b n_{1\alpha} \delta_{\beta\gamma}
	+ c n_{1\beta} \delta_{\alpha\gamma}
	+ d n_{1\gamma} \delta_{\alpha\beta} \\[1ex]
	a
	&= \frac{1}{2\pi} \int g_{\alpha\beta\gamma}
	\bigl(
	4 n_{1\alpha} n_{1\beta} n_{1\gamma}
	- n_{1\alpha}\delta_{\beta\gamma}
	- n_{1\beta}\delta_{\alpha\gamma}\\
	&\hspace{4.26cm}- n_{1\gamma}\delta_{\alpha\beta}
	\bigr)\,
	d\hat{\mathbf{n}}_1d\mathcal{H} \\
	b
	&= \frac{1}{2\pi} \int g_{\alpha\beta\gamma}
	\bigl(
	n_{1\alpha}\delta_{\beta\gamma}
	- n_{1\alpha} n_{1\beta} n_{1\gamma}
	\bigr)\,
	d\hat{\mathbf{n}}_1d\mathcal{H} \\
	c
	&= \frac{1}{2\pi} \int g_{\alpha\beta\gamma}
	\bigl(
	n_{1\beta}\delta_{\alpha\gamma}
	- n_{1\alpha} n_{1\beta} n_{1\gamma}
	\bigr)\,
	d\hat{\mathbf{n}}_1d\mathcal{H} \\
	d
	&= \frac{1}{2\pi} \int g_{\alpha\beta\gamma}
	\bigl(
	n_{1\gamma}\delta_{\alpha\beta}
	- n_{1\alpha} n_{1\beta} n_{1\gamma}
	\bigr)\,
	d\hat{\mathbf{n}}_1d\mathcal{H}.
\end{aligned}
\end{equation}
At four indices things simplify quite a bit. To obtain the equation at fourth order in gradients for the density, the Chapman--Enskog method does not require us to solve for $f^{(4)}$, meaning that after solving for $f^{(3)}$ we just plug the result into the equation at order $\epsilon^4$ and then integrate over $\mathbf{\hat{n}}_1$ to obtain the equation for the density. This implies that we will need to evaluate integrals of the form
\begin{equation}
I_{\alpha\beta\gamma\mu} = \int g_{\alpha\beta\gamma\mu}(\mathbf{\hat{n}}_1,\mathbf{\hat{n}}_2,\bm{\hat{\sigma}})d\hat{\mathbf{n}}_1d\mathcal{H}.
\end{equation}
Since all three unit vectors are integrated out, the result must be an isotropic tensor
\begin{equation}
I_{\alpha\beta\gamma\mu} = a\delta_{\alpha\beta}\delta_{\gamma\mu}+b\delta_{\alpha\gamma}\delta_{\beta\mu}+c\delta_{\alpha\mu}\delta_{\beta\gamma}.
\end{equation}
The constants can then be found by taking all possible permutations of the traces on both sides, that is $\text{Tr}_{\alpha\beta}\text{Tr}_{\gamma\mu}$, $\text{Tr}_{\alpha\gamma}\text{Tr}_{\beta\mu}$ and $\text{Tr}_{\alpha\mu}\text{Tr}_{\beta\gamma}$. Solving the resulting system of equations yields
\begin{equation}\label{eq:I4}
\begin{aligned}
	I_{\alpha\beta\gamma\mu} &= \int g_{\alpha\beta\gamma\mu}(\mathbf{\hat{n}}_1,\mathbf{\hat{n}}_2,\bm{\hat{\sigma}})d\hat{\mathbf{n}}_1d\mathcal{H}\\
	&=a\delta_{\alpha\beta}\delta_{\gamma\mu}+b\delta_{\alpha\gamma}\delta_{\beta\mu}+c\delta_{\alpha\mu}\delta_{\beta\gamma}\\
	a&= \frac{1}{8}\int g_{\alpha\beta\gamma\mu}(3\delta_{\alpha\beta}\delta_{\gamma\mu}-\delta_{\alpha\gamma}\delta_{\beta\mu}-\delta_{\alpha\mu}\delta_{\beta\gamma})d\hat{\mathbf{n}}_1d\mathcal{H}\\
	b&= \frac{1}{8}\int g_{\alpha\beta\gamma\mu}(3\delta_{\alpha\gamma}\delta_{\beta\mu}-\delta_{\alpha\beta}\delta_{\gamma\mu}-\delta_{\alpha\mu}\delta_{\beta\gamma})d\hat{\mathbf{n}}_1d\mathcal{H}\\
	c&= \frac{1}{8}\int g_{\alpha\beta\gamma\mu}(3\delta_{\alpha\mu}\delta_{\beta\gamma}-\delta_{\alpha\beta}\delta_{\gamma\mu}-\delta_{\alpha\gamma}\delta_{\beta\mu})d\hat{\mathbf{n}}_1d\mathcal{H}
\end{aligned}
\end{equation}
\subsection{Chapman--Enskog solution}
As explained in the main text, there are a couple of assumptions needed to be made to apply the Chapman--Enskog method.  First we must assume that the kinetic equation accepts normal solutions, meaning that the spatiotemporal dependence must be enslaved to the conserved fields
\begin{equation}
f(\mathbf{r}_1,\hat{\mathbf{n}}_1,t)=f(\hat{\mathbf{n}}_1,\rho(\mathbf{r}_1,t)).
\end{equation}
When substituting this into the kinetic equation, the time derivative can be written as
\begin{equation}
\frac{\partial f}{\partial t}=\frac{\partial f}{\partial\rho}\frac{\partial\rho}{\partial t}.
\end{equation}
Then, the distribution function is expanded in powers of a small formal parameter $\varepsilon$, that is proportional to each spatial gradient, $f=f^{(0)}+\varepsilon f^{(1)}+\varepsilon^2f^{(2)}+\dots$. Lastly, timescale separation is made explicit by the introduction of several time variables $t_0=t$, $t_1=\varepsilon t$, $t_2=\varepsilon^2 t$ and so forth.  Using these assumptions the distribution function can be written as $f(\mathbf{r}_1,\hat{\mathbf{n}}_1,t)=f[\rho(\mathbf{r}_1,t_0,t_1,t_2,\dots);\hat{\mathbf{n}}_1]$ and the time derivative is expressed using the chain rule
\begin{equation}
\frac{\partial}{\partial t}=\frac{\partial}{\partial t_0}+\varepsilon\frac{\partial}{\partial t_1}+\varepsilon^2\frac{\partial}{\partial t_2}+\dots.
\end{equation}
Now, because the expansion in powers of $\varepsilon$ increases the number of solutions, in the Chapman--Enskog method is customary to introduce the constraint
\begin{equation}\label{eq:normalization}
\int f^{(i)}d\hat{\mathbf{n}}_1=\delta_{0i}\rho,
\end{equation}
i.e., it is demanded that the term of order $\varepsilon^0$ reproduces exactly the hydrodynamic fields.
Introducing these expansions into the kinetic equation yields a hierarchy of equations that can be solved systematically.
\subsubsection{Zeroth order equation}
At order $\epsilon^0$ the kinetic equation reads
\begin{equation}\label{eq:kin0}
\frac{\partial f^{(0)}}{\partial t_0}=D_r \frac{\partial^2 f^{(0)}}{\partial \theta_1^2}.
\end{equation}
By integrating over $\mathbf{\hat{n}}_1$ we obtain for the density 
\begin{equation}\label{eq:rho0}
\frac{\partial \rho}{\partial t_0}=0.
\end{equation}
By substituting the time derivative of $\rho$ back into Eq.~(\ref{eq:kin0}) we obtain a closed equation for $f^{(0)}$
\begin{equation}
\frac{\partial^2 f^{(0)}}{\partial \theta_1^2}=0.
\end{equation}
The periodicity in $\theta$ means that $f^{(0)}$ must be a constant, which upon imposing the normalization condition (\ref{eq:normalization}) gives
\begin{equation}\label{eq:f0}
f^{(0)} = \frac{\rho}{2\pi}.
\end{equation}
\subsubsection{First order equation}
In the next order, $\epsilon^1$, the kinetic equation is
\begin{equation}\label{eq:kin1}
\frac{\partial f^{(1)}}{\partial t_0} + \frac{\partial f^{(0)}}{\partial t_1}+ V\mathbf{\hat{n}}_1\cdot \nabla f^{(0)} = D_r \frac{\partial^2 f^{(1)}}{\partial\theta^2_1} + J^1_{00},
\end{equation}
where we introduce the notation $J^n_{ml}\equiv J^{(n)}[f^{(m)},f^{(l)}]$. Using the result for $f^{(0)}$ and Eqs.~(\ref{eq:J1}) and~(\ref{eq:I1}) give
\begin{equation}
\begin{aligned}
	&J^1_{00}=-\frac{\nabla_\alpha}{4\pi^2}[\chi(\rho)\rho^2]\int\Delta_{\alpha}d\mathcal{H}=-a_1n_{1\alpha}\frac{\nabla_\alpha}{4\pi^2}[\chi(\rho)\rho^2]\\
	&a_1 = \frac{1}{2\pi}\int\Delta_\alpha n_{1\alpha}d\hat{\mathbf{n}}_1d\mathcal{H}=-\frac{1}{2}\pi^2 V\sigma^2
\end{aligned}
\end{equation}
When computing the moment of the kinetic equation at first order [Eq.~(\ref{eq:kin1})], the contribution of the first term vanishes due to the constraint of Eq.~(\ref{eq:normalization}). Furthermore, replacing $f^{(0)}$ with the solution (\ref{eq:f0}) gives
\begin{equation}\label{eq:rho1}
\frac{\partial \rho}{\partial t_1} = 0
\end{equation}
Using (\ref{eq:rho1}) and the explicit form of $f^{(0)}$ results in a closed equation for $f^{(1)}$
\begin{equation}\label{eq:kin11}
\begin{split}
	&\frac{\partial^2 f^{(1)}}{\partial \theta^2_1} = n_{1\alpha}A_\alpha\\
	&A_\alpha = \frac{V}{2\pi D_r}\nabla_\alpha\left[\left(1-\frac{\pi\sigma^2}{4}\chi(\rho)\rho\right)\rho\right]
\end{split}
\end{equation}

This equation can be easily solved by noting that 
\begin{equation}
\frac{\partial^2}{\partial \theta_1^2}n_{1\alpha} = -n_{1\alpha}.
\end{equation}

Then, the solution at the first order is 
\begin{equation}\label{eq:f1}
f^{(1)} = -\frac{n_{1\alpha}}{2\pi D_r}\nabla_\alpha[v(\rho)\rho],
\end{equation}
with $v(\rho)=V[1-\pi\sigma^2\chi(\rho)\rho/4]$.

\subsubsection{Second order equation}
At second order in $\epsilon$, the kinetic equation reads
\begin{equation}\label{eq:kin2}
\frac{\partial f^{(0)}}{\partial t_2}+\frac{\partial f^{(1)}}{\partial t_1}+\frac{\partial f^{(2)}}{\partial t_0}+V\hat{\mathbf{n}}_1\cdot\nabla f^{(1)}= D_r\frac{\partial f^{(0)}}{\partial \theta_1^2}+J_{01}^1+J_{10}^1+J_{00}^2.
\end{equation}
First, we calculate the integrals using Eqs.~(\ref{eq:J2}) and (\ref{eq:I2}) together with the explicit forms of $f^{(0)}$ and $f^{(1)}$ in equations (\ref{eq:f0}) and (\ref{eq:f1})
\begin{equation}\label{eq:J101}
\begin{aligned}
	J^1_{01}&=\frac{\nabla_\alpha}{4\pi^2 D_r}\left\{\chi(\rho)\rho\nabla_\beta[
	v(\rho)\rho]\right\}\int\Delta_\alpha n_{2\beta}d\mathcal{H}\\
	&=\frac{V\sigma^2}{16D_r}\nabla_\alpha\{\chi(\rho)\rho\nabla_\alpha[v(\rho)\rho] \},
\end{aligned}
\end{equation}
\begin{equation}\label{eq:J110}
\begin{aligned}
	J^1_{10}&=\frac{\nabla_\alpha}{4\pi^2D_r}\{\chi(\rho)\rho\nabla_\beta[v(\rho)\rho]
	\}n_{1\beta}\int\Delta_\alpha d\mathcal{H}\\
	&=-\frac{V\sigma^2}{8D_r}\nabla_\alpha\{\chi(\rho)\rho\nabla_\beta[v(\rho)\rho]  \}n_{1\alpha}n_{1\beta},
\end{aligned}
\end{equation}
\begin{widetext}
\begin{equation}\label{eq:J200}
	\begin{aligned}
		J^2_{00}=&\frac{\sigma}{4\pi^2}\nabla_\alpha\{\rho[\chi(\rho)\nabla_\beta\rho+\frac{\rho}{2}\nabla_\beta\chi(\rho)]\}\int\Delta_\alpha \hat{\sigma}_\beta d\mathcal{H}
		+\frac{\nabla_\alpha\nabla_\beta}{8\pi^2}[\chi(\rho)\rho^2]\int \Delta_{\alpha}\Delta_\beta d\mathcal{H},\\
		J^2_{00}=&\frac{V\sigma^2}{8\pi^2}\left[\frac{8}{9}(3C-1)n_{1\alpha}n_{1\beta}+\frac{4}{9}(6C+1)\delta_{\alpha\beta} \right]\nabla_\alpha\left\{2\sigma\rho[\chi(\rho)\nabla_\beta\rho+\frac{\rho}{2}\nabla_\beta\chi(\rho)]+\sigma\nabla_\beta[\chi(\rho)\rho^2]\right\},
	\end{aligned}
\end{equation}
\end{widetext}
where $C\approx0.916$ is the Catalan constant.
Following these results and the expression for $f^{(1)}$ in (\ref{eq:f1}), we can integrate over $\hat{\mathbf{n}}_1$ to obtain a diffusive equation for the density
\begin{equation}\label{eq:rho2}
\frac{\partial \rho}{\partial t_2}  = \nabla_\alpha\left\{D_0\left[1-\frac{(\rho^2\chi)'}{2\rho^*}  \right]\nabla_\alpha\rho \right\},
\end{equation}
where $\rho^*$ is the critical density in the absence of correlations ($\chi=1$), above which a density instability develops, $D_0$ is the bare diffusion coefficient for ABP and the prime denotes derivatives with respect to the density. The explicit form of $\rho^*$ is
\begin{equation}
\rho^* = \frac{2/(\pi\sigma^2)}{1-\frac{16C}{\pi^2\text{Pe}}}.
\end{equation}

To obtain a closed equation for $f^{(2)}$, first we use equation~(\ref{eq:rho2}) in the first term of (\ref{eq:kin2}). The rest of the time derivatives vanish due to Eqs.~(\ref{eq:rho0}) and~(\ref{eq:rho1}) and upon substituting (\ref{eq:J101}-\ref{eq:J200}) along with the expression for $f^{(1)}$ in equation~(\ref{eq:f1}) we obtain
\begin{widetext}
\begin{equation}\label{eq:kin22}
	\begin{aligned}
		D_r\frac{\partial^2 f^{(2)}}{\partial \theta_1^2} &= A[\rho]+B[\rho]_{\alpha\beta}n_{1\alpha}n_{1\beta},\\
		A[\rho] &= \nabla^2\left\{\frac{V^2}{4\pi D_r}\rho-\left[\frac{V^2\sigma^2}{16D_r}+\frac{V\sigma^3(3C-1)}{9\pi^2} \right]\chi\rho^2\right\}-\frac{V\sigma^2}{16D_r}\nabla_\alpha[\chi\rho\nabla_\alpha v(\rho)\rho],\\
		B[\rho]_{\alpha\beta} &= \nabla_\alpha\left\{\frac{V\sigma^2}{8D_r}\chi\rho\nabla_\beta[v(\rho)\rho]-\frac{V}{2\pi D_r}\nabla_\beta[v(\rho)\rho] -\frac{2V\sigma^3(3C-1)}{9\pi^2}\nabla_\beta[\chi\rho^2] \right\}.
	\end{aligned}
\end{equation}

\end{widetext}
Solving Eq.~(\ref{eq:kin22}) is straightforward by using two relations. First, we have that
\begin{equation}\label{eq:dn2}
\frac{\partial^2}{\partial\theta^2} n_{\alpha}n_\beta = -4\left(n_\alpha n_\beta-\frac{\delta_{\alpha\beta}}{2} \right).
\end{equation}
Then, after integrating (\ref{eq:kin22}) over $\theta_1$ both sides must vanish identically by virtue of periodic boundary conditions, meaning that
\begin{equation}
A[\rho]=-\frac{B_{\alpha\alpha}[\rho]}{2}.
\end{equation}
This condition is actually the solvability condition of Eq.~(\ref{eq:kin22}), because the operator $\partial_{\theta\theta}$ is non-invertible. 

The previous condition implies that we can write 
\begin{equation}\label{eq:kin222}
D_r\frac{\partial^2f^{(2)}}{\partial\theta_1^2}=B[\rho]_{\alpha\beta}\left(n_{1\alpha}n_{1\beta}-\frac{\delta_{\alpha\beta}}{2}\right).
\end{equation}
Thus, combining~(\ref{eq:dn2}) and~(\ref{eq:kin222}) gives 
\begin{equation}\label{eq:f2}
\begin{aligned}
	&f^{(2)} =-B[\rho]_{\alpha\beta}\left(n_{1\alpha}n_{1\beta}-\frac{\delta_{\alpha\beta}}{2}\right),\\
	&B[\rho]_{\alpha\beta}=\{A_1\nabla_\alpha\nabla_\beta[v(\rho)\rho]+A_2\nabla_\alpha\rho\nabla_\beta(\chi\rho^2)\\
	&\hspace{3.5cm}-A_3\nabla_\alpha(\chi\rho\nabla_\beta[v(\rho)\rho]) \},\\
	&A_1 = \frac{V}{8\pi D_r^2},\quad A_2=\frac{V\sigma^3(3C-1)}{18\pi^2D_r},\quad A_3= \frac{V\sigma^2}{32 D_r^2}.
\end{aligned}
\end{equation}

\subsubsection{Third order equation}
At order $\epsilon^3$ the kinetic equation is 
\begin{multline}\label{eq:kin3}
\frac{\partial f^{(0)}}{\partial t_3}+\frac{\partial f^{(2)}}{\partial t_1}+\frac{\partial f^{(1)}}{\partial t_2}+\frac{\partial f^{(3)}}{\partial t_0}+V\nabla_\alpha n_{1\alpha}f^{(2)}\\=D_r\frac{\partial^2 f^{(3)}}{\partial\theta_1^2}+J^1_{02}+J^1_{20}+J^1_{11}+J^2_{01}+J^2_{10}+J^3_{00}.
\end{multline}
As before, we first proceed by calculating the integral terms appearing in the equation, by using (\ref{eq:J3}) and (\ref{eq:I3}) together with the solutions to $f^{(0-2)}$, resulting in
\begin{equation}\label{eq:J102}
\begin{aligned}
	J^1_{02}&=\frac{\nabla_\alpha}{2\pi}(\chi\rho B[\rho]_{\beta\gamma})\int \bm{\Delta}_{\alpha}\left(n_{2\beta}n_{2\gamma}-\frac{\delta_{\beta\gamma}}{2}\right)d\mathcal{H}\\
	&=0,
\end{aligned}
\end{equation}
\begin{equation}\label{eq:J120}
\begin{aligned}
	J^1_{20}  &=\frac{\nabla_\alpha}{2\pi}(\chi\rho B[\rho]_{\beta\gamma})\left(n_{1\beta}n_{1\gamma}-\frac{\delta_{\beta\gamma}}{2}\right)\int\bm{\Delta}_\alpha d\mathcal{H}\\
	&=-\frac{V\sigma^2\pi}{2}\nabla_\alpha(\chi\rho B[\rho]_{\beta\gamma})n_{1\alpha}\left(n_{1\beta}n_{1\gamma}-\frac{\delta_{\beta\gamma}}{2}\right),
\end{aligned}
\end{equation}
\begin{equation}\label{eq:J111}
\begin{aligned}
	J^1_{11}&=-\frac{\nabla_\alpha}{4\pi^2D_r^2}\{\chi\nabla_\beta[v(\rho)\rho]\nabla_\gamma[v(\rho)\rho] \}n_{1\beta}\int\bm{\Delta}_\alpha n_{2\gamma}d\mathcal{H}\\
	&=-\frac{V\sigma^2}{16D_r^2}\nabla_\alpha\{\chi\nabla_\beta[v(\rho)\rho]\nabla_\gamma[v(\rho)\rho] \}n_{1\beta}\delta_{\alpha\gamma},
\end{aligned}
\end{equation}
\begin{widetext}
\begin{equation}\label{eq:J210a}
	\begin{aligned}
		J^2_{10}&=\frac{\sigma}{4\pi^2D_r}\nabla_\alpha\{\chi_{1\beta}[\rho]\nabla_\gamma[v(\rho)\rho] \}n_{1\gamma}\int\bm{\Delta}_\alpha\hat{\sigma}_\beta d\mathcal{H}-\frac{\nabla_\alpha\nabla_\beta}{8\pi^2D_r}\{\chi\rho\nabla_\gamma[v(\rho)\rho]\}n_{1\gamma}\int\bm{\Delta}_\alpha\bm{\Delta}_\beta d\mathcal{H}\\
		&=-\frac{V\sigma^3}{18\pi^2D_r}[2(3C-1)n_{1\alpha}n_{1\beta}+(6C+1)\delta_{\alpha\beta}]n_{1\gamma}\nabla_\alpha\{2\chi_{1\beta}[\rho]\nabla_\gamma[v(\rho)\rho]+\nabla_\beta[\chi\rho\nabla_\gamma(v(\rho)\rho)] \},
	\end{aligned}
\end{equation}
where we have defined $\chi_{1\beta}[\rho]=\chi\nabla_\beta\rho+\rho\nabla_\beta\chi/2$,
\begin{equation}\label{eq:J210b}
	\begin{aligned}
		J^2_{01}&=\frac{\sigma}{4\pi^2D_r}\nabla_\alpha\{\rho\chi_{1\beta}[\nabla_\gamma( v(\rho)\rho)] \}\int\bm{\Delta}_\alpha\hat{\sigma}_\beta n_{2\gamma} d\mathcal{H}-\frac{\nabla_\alpha\nabla_\beta}{8\pi^2D_r}\{\chi\rho\nabla_\gamma[v(\rho)\rho]\}\int\bm{\Delta}_\alpha\bm{\Delta}_\beta n_{2\gamma}d\mathcal{H}\\
		&=\frac{V\sigma^3}{90\pi^2D_r}[-2(3C-1)n_{1\alpha}n_{1\beta}n_{1\gamma}+8(3C-1)(n_{1\alpha}\delta_{\beta\gamma}+n_{1\beta}\delta_{\alpha\gamma})+(7-6C)n_{1\gamma}\delta_{\alpha\beta} ]\\
		&\qquad\qquad\times\nabla_\alpha\{2\rho\chi_{1\beta}[\nabla_\gamma(v(\rho)\rho)]+\nabla_\beta(\chi\rho\nabla_\gamma[v(\rho)\rho]) \},
	\end{aligned}
\end{equation}
\begin{equation}
	\begin{aligned}
		J^3_{00}&=-\frac{\sigma^2}{8\pi^2} \nabla_\alpha(\rho\chi_{2\beta\gamma}[\rho])\int\Delta_\alpha\hat{\sigma}_\beta\hat{\sigma}_\gamma d\mathcal{H}+\frac{\sigma}{8\pi^2}\nabla_\alpha\nabla_\beta(\rho\chi_{1\gamma}[\rho])\int\Delta_\alpha \Delta_\beta \hat{\sigma}_\gamma d\mathcal{H}-\frac{\nabla_\alpha\nabla_\beta\nabla_\gamma}{24\pi^2}(\chi\rho^2)\int\Delta_\alpha\Delta_\beta\Delta_\gamma d\mathcal{H}\\
		&=-\frac{V\sigma^4}{192\pi}[(5\pi-8)n_{1\alpha}\delta_{\beta\gamma}-(8+\pi)(n_{1\beta}\delta_{\alpha\gamma}+n_{1\gamma}\delta_{\alpha\beta})]\nabla_\alpha(\rho\chi_{2\beta\gamma}[\rho])\\
		&\hspace{3cm}+\frac{V\sigma^4}{1536\pi}[3(16-8\pi+\pi^3)(n_{1\alpha}\delta_{\beta\gamma}+n_{1\beta}\delta_{\alpha\gamma})+(40\pi+3\pi^3-112)n_{1\gamma}\delta_{\alpha\beta}\nabla_\alpha\nabla_\beta(\rho\chi_{1\gamma}[\rho])\\
		&\hspace{6.25cm}+\frac{V\sigma^4(9\pi^3-8\pi-16)}{9216\pi}(n_{1\alpha}\delta_{\beta\gamma}+n_{1\beta}\delta_{\alpha\gamma}+n_{1\gamma}\delta_{\alpha\beta})\nabla_\alpha\nabla_\beta\nabla_\gamma(\chi\rho^2),
	\end{aligned}
\end{equation}
\end{widetext}
with $\chi_{2\beta\gamma}[f]=\chi\nabla_\beta\nabla_\gamma f +\nabla_\beta\chi\nabla_\gamma f+\frac{f}{4}\nabla_\beta\nabla_\gamma\chi$.

Because the density is a scalar field, it is easy to see that at this time scale $\partial_{t_3}\rho=0$, meaning that the only task left is to solve for $f^{(3)}$. The process is similar to the one used to find $f^{(2)}$ but it is a lot more tedious due to the long expressions found at this order. Because of the normalization condition (\ref{eq:normalization}) and that the density is stationary at this time scale, the first four terms in equation~(\ref{eq:kin3}) vanish. Now, similarly to the previous orders, we will want to write the equation for $f^{(3)}$ in the form
\begin{equation}\label{eq:kin33}
D_r\frac{\partial f^{(3)}}{\partial\theta_1^2}=C_\alpha n_{1\alpha}+D_{\alpha\beta\gamma}n_{1\alpha}n_{1\beta}n_{1\gamma},
\end{equation}
where it is understood that $\mathbf{C}=\mathbf{C}[\rho]$ and $\mathbb{D}=\mathbb{D}[\rho]$.
In this case, because we have an odd number of $\hat{\mathbf{n}}$, the right hand side automatically satisfies the solvability condition and hence there is no relation to be imposed between $C_\alpha$ and $D_{\alpha\beta\gamma}$. To proceed we must first note that
\begin{equation}\label{eq:nnn}
\frac{\partial^2 }{\partial\theta^2}n_{\alpha}n_{\beta}n_{\gamma}=-9\left[n_\alpha n_\beta n_\gamma -\frac{2}{9}(n_\alpha\delta_{\beta\gamma}+n_\beta\delta_{\alpha\gamma}+n_\gamma\delta_{\alpha\beta})\right].
\end{equation}
Then, we propose a solution of the form
\begin{equation}\label{eq:f3ans}
f^{(3)}= C'_\alpha n_{1\alpha}+D'_{\alpha\beta\gamma}n_{1\alpha}n_{1\beta}n_{1\gamma}
\end{equation}
and after applying $D_r\partial_{\theta_1\theta_1}$ it is found that
\begin{equation}
\begin{aligned}
	D'_{\alpha\beta\gamma}&=-\frac{D_{\alpha\beta\gamma}}{9D_r},\\
	C'_\alpha &= -\left[C_\alpha +\frac{2}{9}(D_{\alpha\beta\beta}+D_{\beta\alpha\beta}+D_{\beta\beta\alpha})\right].
\end{aligned}
\end{equation}
With this, the explicit solution to $f^{(3)}$ is
\begin{multline}\label{eq:f3}
f^{(3)}= -\frac{1}{D_r}\left[ C_\alpha+\frac{2}{9}(D_{\alpha\beta\beta}+D_{\beta\alpha\beta}+D_{\beta\beta\alpha})\right]n_{1\alpha}\\-\frac{D_{\alpha\beta\gamma}}{9D_r}n_{1\alpha}n_{1\beta}n_{1\gamma}.
\end{multline}
Finally, at this order, it remains to find explicit expressions for $C_\alpha$ and $D_{\alpha\beta\gamma}$. The process is straightforward but cumbersome and it only consists of making the contractions on all the integral terms on the right hand side of (\ref{eq:kin3}) and on the fifth term on the left hand side. Here we present the final result
\begin{widetext}
\begin{multline}
	C_\alpha= \frac{V}{2}\nabla_\alpha B_{\beta\beta}+\frac{V\pi\sigma^2}{8}\nabla_\alpha(\chi\rho B_{\beta\beta})+\frac{V\sigma^2}{16D_r^2}\nabla_\beta\{\chi\nabla_\alpha[v(\rho)\rho]\nabla_\beta[(v(\rho)\rho)] \}\\-\frac{8V\sigma^3(3C-1)}{45\pi^2D_r}\{\nabla_\alpha(\rho\chi_{1\beta}[\nabla_\beta( v(\rho)\rho)])+\nabla_\beta(\rho\chi_{1\alpha}[\nabla_\beta(v(\rho)\rho)])+\nabla_\alpha\nabla_\beta(\chi\rho\nabla_\beta[v(\rho)\rho]) \}\\-\frac{V\sigma^3(7-6C)}{90\pi^2D_r}\nabla_\beta(\rho\chi_{1\beta}[\nabla_\alpha(v(\rho)\rho)])+\frac{V\sigma^3(6C+1)}{9\pi^2D_r}\nabla_\beta\{\chi_{1\beta}[\rho]\nabla_\alpha[v(\rho)\rho] \}+\frac{V\sigma^3(18C-1)}{45\pi^2D_r}\nabla_\beta\nabla_\beta\{\chi\rho\nabla_\alpha[v(\rho)\rho]\}\\-\frac{V\sigma^4(8+\pi)}{96\pi}\nabla_\beta(\rho\chi_{2\alpha\beta}[\rho])-\frac{V\sigma^4(16-8\pi+\pi^3)}{256}\nabla_\alpha\nabla_\beta(\rho\chi_{1\beta}[\rho])-\frac{V\sigma^4(40\pi+3\pi^2-112)}{1536\pi}\nabla_\beta\nabla_\beta(\rho\chi_{1\alpha}[\rho])\\-\frac{V\sigma^4(9\pi^3-8\pi-16)}{3072}\nabla_\alpha\nabla_\beta\nabla_\beta(\chi\rho^2)+\frac{V\sigma^4(5\pi-8)}{192\pi}\nabla_\alpha(\rho\chi_{2\beta\beta}[\rho]),
\end{multline}

\begin{equation}
	D_{\alpha\beta\gamma}=-\nabla_\alpha[v(\rho)B_{\beta\gamma}]+
	\frac{2V\sigma^3(3C-1)}{45\pi^2D_r}\nabla_\alpha\{\rho\chi_{1\beta}[\nabla_\gamma(v(\rho)\rho)]+5\chi_{1\beta}[\rho]\nabla_\gamma[v(\rho)\rho]+3\nabla_\beta(\chi\rho\nabla_\gamma[v(\rho)\rho])\}.
\end{equation}
\end{widetext}

\subsubsection{Fourth order equation}
At the fourth order, we need to directly integrate the kinetic equation to obtain an equation for the density. Thus, after integrating over $\hat{\mathbf{n}}$ we are left with 
\begin{multline}
\frac{\partial \rho}{\partial t_4}=-V\nabla_\alpha\int n_{1\alpha}f^{(3)}d\hat{\mathbf{n}}_1+ I^1_{03}+I^1_{30}+I^1_{12}+I^1_{21}\\+I^2_{02}+I^2_{20}+I^2_{11}+I^3_{01}+I^3_{10}+I^4_{00},
\end{multline}
where we used the normalization condition (\ref{eq:normalization}) to cancel out the other time derivatives and the terms $I^n_{ml}$ are simply the integrals of $J^n_{ml}$ over $\hat{\mathbf{n}}$. Similar to the previous orders, we start by calculating the integral terms for which we need the forms of $f^{(0-3)}$ in Eqs.~(\ref{eq:f0}),~(\ref{eq:f1}),~(\ref{eq:f2}) and~(\ref{eq:f3}). For brevity, even though we have an explicit form for $f^{(3)}$, for the rest of the calculations we will write it as in equation~(\ref{eq:f3ans}). Now, as shown in section~\ref{sect:integrals}, because we are integrating over all unit vectors, the results of the integrals are linear combinations of isotropic tensors. Then, by (\ref{eq:I4}) we have
\begin{multline*}
I^1_{03}=-\frac{\nabla_\alpha}{2\pi}\left(\chi\rho C'_\beta\int \Delta_\alpha n_{2\beta}d\hat{\mathbf{n}}_1d\mathcal{H} \right.\\\left.+\chi\rho D'_{\beta\gamma\mu}\int\Delta_\alpha n_{2\gamma}n_{2\beta}n_{2\mu}d\hat{\mathbf{n}}_1d\mathcal{H} \right),
\end{multline*}
\begin{multline}
I^1_{03}=-\frac{V\pi^2\sigma^2}{16}\nabla_\alpha\{\chi\rho[4C'_\alpha+(D'_{\alpha\beta\beta}+D'_{\beta\alpha\beta}\\+D'_{\beta\beta\alpha})]\},
\end{multline}
\begin{multline*}
I^1_{30}=-\frac{\nabla_\alpha}{2\pi}\left(\chi\rho C'_\beta\int \Delta_\alpha n_{1\beta}d\hat{\mathbf{n}}_1d\mathcal{H} \right.\\\left.+\chi\rho D'_{\beta\gamma\mu}\int\Delta_\alpha n_{1\gamma}n_{1\beta}n_{1\mu}d\hat{\mathbf{n}}_1d\mathcal{H} \right),
\end{multline*}
\begin{multline}
I^1_{30}=\frac{V\pi^2\sigma^2}{16}\nabla_\alpha\left\{\chi\rho\left[4C'_\alpha+(D'_{\alpha\beta\beta}+D'_{\beta\alpha\beta}+D'_{\beta\beta\alpha})\right]\right\},
\end{multline}
\begin{multline*}
I^1_{12}=-\frac{\nabla_\alpha}{2\pi D_r}\{\chi B_{\gamma\mu}\nabla_\beta[v(\rho)\rho]\}\\\times \int\Delta_\alpha n_{1\beta}\left(n_{2\gamma}n_{2\mu}-\frac{\delta_{\gamma\mu}}{2}\right)d\hat{\mathbf{n}}_1d\mathcal{H}
\end{multline*}
\begin{multline}
I^1_{12}= -\frac{V\pi^2\sigma^2}{16 D_r}\nabla_\alpha\{\chi B_{\gamma\mu}\nabla_\beta[v(\rho)\rho]\}(\delta_{\alpha\gamma}\delta_{\beta\mu}\\+\delta_{\alpha\mu}\delta_{\beta\gamma}-\delta_{\alpha\beta}\delta_{\gamma\mu}),
\end{multline}
\begin{multline*}
I^1_{21}=-\frac{\nabla_\alpha}{2\pi D_r}\{\chi B_{\beta\gamma}\nabla_\mu[v(\rho)\rho]\}\\\times \int\Delta_\alpha n_{2\mu}\left(n_{1\beta}n_{1\gamma}-\frac{\delta_{\beta\gamma}}{2}\right)d\hat{\mathbf{n}}_1d\mathcal{H},
\end{multline*}
\begin{multline}
I^1_{21}=\frac{V\pi^2\sigma^2}{16D_r}\nabla_\alpha\{\chi B_{\beta\gamma}\nabla_\mu[v(\rho)\rho]\}(\delta_{\alpha\beta}\delta_{\gamma\mu}\\-\delta_{\alpha\gamma}\delta_{\beta\mu}-\delta_{\alpha\mu}\delta_{\beta\gamma}).
\end{multline}
In the previous results, we have $I^1_{03}+I^1_{30}=0$ and $I^1_{21}+I^1_{12}=0$, where the last equality can be obtained by a simple renaming of the indices, $\beta\rightarrow\gamma$, $\gamma\rightarrow\mu$ and $\mu\rightarrow\beta$. The fact that these integrals cancel each other out comes from a parity symmetry in the angular integrals when changing $\hat{\mathbf{n}}_1\leftrightarrow\hat{\mathbf{n}}_2$ which can probably be proven using the explicit expression for the displacement $\bm{\Delta}$. The proof falls out of the scope of this article so we leave it for any interested reader. 

For the 6 remaining integrals:
\begin{widetext}
\begin{equation}\label{eq:I202}
	\begin{aligned}
		I^2_{02}&=\frac{\sigma}{2\pi}\nabla_\alpha(\rho\chi_{1\beta} [B_{\gamma\mu}]\int \Delta_\alpha\hat{\sigma}_\beta\left(n_{2\gamma}n_{2\mu}-\frac{\delta_{\gamma\mu}}{2}\right)d\hat{\mathbf{n}}_1d\mathcal{H}-\frac{\nabla_\alpha\nabla_\beta}{4\pi}(\chi \rho B_{\gamma\mu})\int\Delta_\alpha\Delta_\beta\left(n_{2\gamma}n_{2\mu}-\frac{\delta_{\gamma\mu}}{2} \right)d\hat{\mathbf{n}}_1d\mathcal{H}\\
		&=-\frac{V\sigma^3(3C-1)}{18}(\delta_{\alpha\gamma}\delta_{\beta\mu}+\delta_{\alpha\mu}\delta_{\beta\gamma}-\delta_{\alpha\beta}\delta_{\gamma\mu})\nabla_\alpha\{2\rho\chi_{1\beta} [B_{\gamma\mu}]+\nabla_\beta(\chi\rho B_{\gamma\mu})\},
	\end{aligned}
\end{equation}
\begin{equation}\label{eq:I220}
	\begin{aligned}
		I^2_{20}&=\frac{\sigma}{2\pi}\nabla_\alpha(\chi_{1\beta}[\rho]B_{\gamma\mu})\int \Delta_\alpha\hat{\sigma}_\beta\left(n_{1\gamma}n_{1\mu}-\frac{\delta_{\gamma\mu}}{2}\right)d\hat{\mathbf{n}}_1d\mathcal{H}-\frac{\nabla_\alpha\nabla_\beta}{4\pi}(\chi\rho B_{\gamma\mu})\int\Delta_\alpha\Delta_\beta\left(n_{1\gamma}n_{1\mu}-\frac{\delta_{\gamma\mu}}{2} \right)d\hat{\mathbf{n}}_1d\mathcal{H}\\
		&=-\frac{V\sigma^3(3C-1)}{18}(\delta_{\alpha\gamma}\delta_{\beta\mu}+\delta_{\alpha\mu}\delta_{\beta\gamma}-\delta_{\alpha\beta}\delta_{\gamma\mu})\nabla_\alpha\{2\chi_{1\beta} [\rho]B_{\gamma\mu}+\nabla_\beta(\chi\rho B_{\gamma\mu})\},
	\end{aligned}
\end{equation}
\begin{multline*}
	I^2_{11}=  -\frac{\sigma}{4\pi^2D_r^2}\{\chi_{1\beta}[\nabla_\mu(v(\rho)\rho)]\nabla_\gamma[v(\rho)\rho]\}\int\Delta_\alpha\hat{\sigma}_\beta n_{1\gamma}n_{2\mu}d\hat{\mathbf{n}}_1d\mathcal{H}\\
	+\frac{\nabla_\alpha\nabla_\beta}{8\pi^2 D_r}\{\chi\nabla_\gamma[v(\rho)\rho]\nabla_\mu[v(\rho)\rho])\int \Delta_\alpha\Delta_\beta n_{1\gamma}n_{2\mu}d\hat{\mathbf{n}}_1d\mathcal{H}
\end{multline*}
\begin{multline}\label{eq:I211}
	\phantom{I^2_{11}}=-\frac{V\sigma^2}{12\pi D_r^2}[(1-C)\delta_{\alpha\beta}{\delta_\gamma\mu}+(3C-1)(\delta_{\alpha\gamma}\delta_{\beta\mu}+\delta_{\alpha\mu}\delta_{\beta\gamma})]\nabla_\alpha\{2\nabla_\gamma[v(\rho)\rho]\chi_{1\beta}[\nabla_\mu(v(\rho)\rho)]+\nabla_\beta(\chi\nabla_\gamma[v(\rho)\rho]\nabla_\mu[v(\rho)\rho])\},
\end{multline}
\begin{multline*}
	I^3_{01}=\frac{\sigma^2}{8\pi^2D_r}\nabla_\alpha\{\rho\chi_{2\beta\gamma}[\nabla_\mu(v(\rho)\rho)]\}\int\Delta_\alpha\hat{\sigma}_\beta\hat{\sigma}_\gamma n_{2\mu}d\hat{\mathbf{n}}_1d\mathcal{H}-\frac{\sigma}{8\pi^2D_r}\nabla_\alpha\nabla_\beta\{\rho\chi_{1\gamma}[\nabla_\mu(v(\rho)\rho)]\}\int\Delta_\alpha\Delta_\beta\hat{\sigma}_\gamma n_{2\mu}d\hat{\mathbf{n}}_1d\mathcal{H}\\+\frac{\nabla_\alpha\nabla_\beta\nabla_\gamma}{24\pi^2D_r}\{\chi\rho\nabla_\mu[v(\rho)\rho]\}\int\Delta_\alpha\Delta_\beta\Delta_\gamma n_{2\mu}d\hat{\mathbf{n}}_1d\mathcal{H}
\end{multline*}
\begin{multline}\label{eq:I301}
	\phantom{I^3_{01}}=  \frac{V\sigma^4}{192D_r}[(8+\pi)(\delta_{\alpha\beta}\delta_{\gamma\mu}+\delta_{\alpha\gamma}\delta_{\beta\mu})+(5\pi-8)\delta_{\alpha\mu}\delta_{\beta\gamma}]\nabla_\alpha\{\rho\chi_{2\beta\gamma}[\nabla_\mu(v(\rho)\rho)]\}\\+\frac{V\sigma^4}{1536D_r}[(3\pi^3+40\pi-112)\delta_{\alpha\beta}\delta_{\gamma\mu}+3(16+\pi^3-8\pi)(\delta_{\alpha\gamma}\delta_{\beta\mu}+\delta_{\alpha\mu}\delta_{\beta\gamma})]\nabla_\alpha\nabla_\beta(\rho\chi_{1\gamma}[\nabla_\mu(v(\rho)\rho)])\\+\frac{V\sigma^4}{9216D_r}(9\pi^3-8\pi-16)(\delta_{\alpha\beta}\delta_{\gamma\mu}+\delta_{\alpha\gamma}\delta_{\beta\mu}+\delta_{\alpha\mu}\delta_{\beta\gamma})\nabla_\alpha\nabla_\beta\nabla_\gamma\{\chi\rho\nabla_\mu[v(\rho)\rho]\},
\end{multline}

\begin{multline*}
	I^3_{10}=\frac{\sigma^2}{8\pi^2D_r}\nabla_\alpha\{\chi_{2\beta\gamma}[\rho]\nabla_\mu[v(\rho)\rho]\}\int\Delta_\alpha\hat{\sigma}_\beta\hat{\sigma}_\gamma n_{1\mu}d\hat{\mathbf{n}}_1d\mathcal{H}-\frac{\sigma}{8\pi^2D_r}\nabla_\alpha\nabla_\beta\{\chi_{1\gamma}[\rho]\nabla_\mu[v(\rho)\rho]\}\int\Delta_\alpha\Delta_\beta\hat{\sigma}_\gamma n_{1\mu}d\hat{\mathbf{n}}_1d\mathcal{H}\\+\frac{\nabla_\alpha\nabla_\beta\nabla_\gamma}{24\pi^2D_r}\{\chi\rho\nabla_\mu[v(\rho)\rho]\}\int\Delta_\alpha\Delta_\beta\Delta_\gamma n_{1\mu}d\hat{\mathbf{n}}_1d\mathcal{H}
\end{multline*}
\begin{multline}\label{eq:I310}
	\phantom{I^3_{10}}=  -\frac{V\sigma^4}{192D_r}[(8+\pi)(\delta_{\alpha\beta}\delta_{\gamma\mu}+\delta_{\alpha\gamma}\delta_{\beta\mu})+(5\pi-8)\delta_{\alpha\mu}\delta_{\beta\gamma}]\nabla_\alpha\{\chi_{2\beta\gamma}[\rho]\nabla_\mu[v(\rho)\rho]\}\\-\frac{V\sigma^4}{1536D_r}[(3\pi^3+40\pi-112)\delta_{\alpha\beta}\delta_{\gamma\mu}+3(16+\pi^3-8\pi)(\delta_{\alpha\gamma}\delta_{\beta\mu}+\delta_{\alpha\mu}\delta_{\beta\gamma})]\nabla_\alpha\nabla_\beta\{\chi_{1\gamma}[\rho]\nabla_\mu[v(\rho)\rho]\}\\-\frac{V\sigma^4}{9216D_r}(9\pi^3-8\pi-16)(\delta_{\alpha\beta}\delta_{\gamma\mu}+\delta_{\alpha\gamma}\delta_{\beta\mu}+\delta_{\alpha\mu}\delta_{\beta\gamma})\nabla_\alpha\nabla_\beta\nabla_\gamma\{\chi\rho\nabla_\mu[v(\rho)\rho]\},
\end{multline}
\begin{multline*}
	I^4_{00}=-\frac{\sigma^3}{24\pi^2}\nabla_\alpha(\rho\chi_{3\beta\gamma\mu}[\rho])\int\Delta_\alpha\hat{\sigma}_\beta\hat{\sigma}_\gamma\hat{\sigma}_\mu d\hat{\mathbf{n}}_1d\mathcal{H}+\frac{\sigma^2}{16\pi^2}\nabla_\alpha\nabla_\beta(\rho\chi_{2\gamma\mu}[\rho])\int\Delta_\alpha\Delta_\beta\hat{\sigma}_\gamma\hat{\sigma}_\mu d\hat{\mathbf{n}}_1d\mathcal{H}\\
	-\frac{\sigma}{24\pi^2}\nabla_\alpha\nabla_\beta\nabla_\gamma(\rho\chi_{1\mu}[\rho])\int\Delta_\alpha\Delta_\beta\Delta_\gamma\hat{\sigma}_\mu d\hat{\mathbf{n}}_1d\mathcal{H}+\frac{\nabla_\alpha\nabla_\beta\nabla_\gamma\nabla_\mu}{96\pi^2}(\chi\rho^2)\int\Delta_\alpha\Delta_\beta\Delta_\gamma\Delta_\mu d\hat{\mathbf{n}}_1d\mathcal{H},
\end{multline*}
\begin{multline}
	\phantom{I^4_{00}}=\frac{V\sigma^5C}{12\pi}(\delta_{\alpha\beta}\delta_{\gamma\mu}+\delta_{\alpha\gamma}\delta_{\beta\mu}+\delta_{\alpha\mu}\delta_{\beta\gamma})\nabla_\alpha(\rho\chi_{3\beta\gamma\mu}[\rho])-\frac{V\sigma^5}{36\pi}[(5C-\text{ln}~4)\delta_{\alpha\beta}\delta_{\gamma\mu}+(C+\text{ln}~4)](\delta_{\alpha\gamma}\delta_{\beta\mu}+\delta_{\alpha\mu}\delta_{\beta\gamma})\nabla_\alpha\nabla_\beta(\rho\chi_{2\gamma\mu}[\rho])\\-\frac{V\sigma^5}{288\pi}[2C+\text{ln}~2-9\text{i~Li}_4(-\text{i})+9\text{i~Li}_4(\text{i})](\delta_{\alpha\beta}\delta_{\gamma\mu}+\delta_{\alpha\gamma}\delta_{\beta\mu}+\delta_{\alpha\mu}\delta_{\beta\gamma})\nabla_\alpha\nabla_\beta\nabla_\gamma\{2\rho\chi_{1\mu}[\rho]+\nabla_\mu(\chi\rho^2)\},
\end{multline}
\end{widetext}

with $\text{Li}_s(z)$ the polylogarithm function of order $s$ and 
\begin{multline}
\chi_{3\beta\gamma\mu}[f]=\frac{\chi}{3!}\nabla_\beta\nabla_\gamma\nabla_\mu f+\frac{1}{4}\nabla_\beta\chi\nabla_\gamma\nabla_\mu f\\+\frac{1}{8}\nabla_\beta\nabla_\gamma\chi\nabla_\mu f+\frac{f}{8\cdot 3!}\nabla_\beta\nabla_\gamma\nabla_\mu\chi . 
\end{multline}

With all of the integrals calculated, it remains to make the contractions of the Kronecker deltas and simplify the resulting equation for the density at order $\varepsilon^4$ as much as possible. This yields all terms of $\mathbf{G}[\rho]$ mentioned in the main text. Finally, the total equation for the density is obtained by summing the dynamics at all orders by replacing $t_n=\varepsilon^nt$ and then $\varepsilon$ is set to unity.

\end{document}